\documentclass[12pt]{iopart}
\usepackage{graphicx}
\usepackage{amsmath}
\usepackage{iopams}
\usepackage{bm}

\begin{document}

\title{Entropy estimates of small data sets}
\author{Juan A. Bonachela$^{1,2}$, Haye Hinrichsen$^2$, and
Miguel A. Mu\~noz$^1$}

\address{$^1$ Departamento de Electromagnetismo y F{\'\i}sica de la Materia and \\
Instituto de F{\'\i}sica Te{\'o}rica y Computacional Carlos I,\\
Facultad de Ciencias, Universidad de Granada, 18071 Granada, Spain\\
$^2$ Fakult\"at f\"ur Physik und Astronomie, Universit\"at W\"urzburg,\\
Am Hubland, 97074 W\"urzburg, Germany} 


\begin{abstract}
  Estimating entropies from limited data series is known to be a
  non-trivial task. Na\"ive estimations are plagued with both
  systematic (bias) and statistical errors. Here, we present a new
  ``balanced estimator'' for entropy functionals (Shannon, R\'enyi and
  Tsallis) specially devised to provide a compromise between
  low bias and small statistical errors, for short data series.  This
  new estimator out-performs other currently available ones when the
  data sets are small and the probabilities of the possible outputs of
  the random variable are not close to zero. Otherwise, other
  well-known estimators remain a better choice. The potential range of
  applicability of this estimator is quite broad specially for
  biological and digital data series.
\end{abstract}

\vspace{2pc}
\noindent{ Journal Reference: J. Phys. A: Math. Theor. 41 (2008) 202001.}

\vspace{0.5cm}
\section{Introduction}

In statistical mechanics and information theory, entropy is a
functional that measures the information content of a statistical
ensemble or equivalently the uncertainty of a random variable. Its
applications in physics, biology, computer science, linguistic, etc
are countless. For example, it has become a key tool in data mining
tasks arising from high-throughput biological analyses.

Historically, the most important example of such a functional is the
Shannon (or {\it information}) entropy \cite{Shannon,Cover}.  For a
discrete random variable $x$, which can take a finite number, $M$, of
possible values $x_{i}\in\left\lbrace x_{1},\hdots,
  x_{M}\right\rbrace$ with corresponding probabilities
$p_{i}\in\left\lbrace p_{1},\hdots, p_{M}\right\rbrace$, this entropy
is defined by:
\begin{equation}
\begin{array}{rl}
H_{S}=&-\displaystyle \sum_{i=1}^{M}p_{i}\textnormal{ln}(p_{i}).
\end{array}
\label{shannon_def}
\end{equation}

Recently, various generalizations, inspired by the study of
$q$-deformed algebras and special functions, have been investigated,
most notably the R\'enyi entropy \cite{Renyi}:
\begin{equation}
\begin{array}{rl}
 H_{R}(q)=&\dfrac{1}{1-q}\textnormal{ln}\left(\displaystyle \sum_{i=1}^{M}p^{q}_{i}\right),
\end{array}
\label{renyi_def}
\end{equation}
with $p \geq 0$, which, in particular, reduces to the Shannon entropy
in the limit $q\rightarrow 1$.
Also, the Tsallis entropy \cite{Tsallis}:
\begin{equation}
\begin{array}{rl}
H_{T}(q)=&\dfrac{1}{q-1}\left( 1-\displaystyle \sum_{i=1}^{M}p^{q}_{i}\right),
\end{array}
\label{tsallis_def}
\end{equation}
although controversial, has generated a large burst of research activity.

In general, the full probability distribution for a given stochastic
problem is not known and, in particular, in many situations only small
data sets from which to infer entropies are available. For example, it
could be of interest to determine the Shannon entropy of a given DNA
sequence.  In such a case, one could {\it estimate} the probability of
each element $i$ to occur, $p_i$, by making some assumption on the
probability distribution, as for example (i) parametrizing it
\cite{param}, (ii) dropping the most unlikely values
\cite{truncation_seq} or (iii) assuming some {\it a priori} shape for
the probability distribution~\cite{bayesian,Wolpert}.  However, the
easiest and most objective way to estimate them is just by counting
how often the value $x_{i}$ appears in the data set
\cite{Grass1,Grass2,Schuermann, Roulston,Miller,Harris,Herzel}.
Denoting this number by $n_{i}$ and dividing by the total size of the
data set one obtains the relative frequency:

\begin{equation}
\begin{array}{rl}
  \hat{p}_{i}=&\dfrac{n_{i}}{N}
\end{array}
\label{frec}
\end{equation}
which {\it na\"ively} approximates the probability $p_{i}$. Obviously,
the entropy of the data set can be approximated by simply replacing
the probabilities $p_{i}$ by $\hat{p}_{i}$ in the entropy
functional.
For example, the Shannon entropy can be estimated by:
\begin{equation}
\begin{array}{rl}
  H_{S}\approx \hat{H}_{S}^{naive}=&-\displaystyle \sum_{i=1}^{M}\hat{p}_{i}
  \textnormal{ln}(\hat{p}_{i})=-\displaystyle \sum_{i=1}^{M}\dfrac{n_{i}}{N}
  \textnormal{ln}\left( \dfrac{n_{i}}{N}\right). 
\end{array}
\label{shannon_naive}
\end{equation}
The quantity $\hat{H}_{S}^{naive}$ is an example of an {\it estimator}
of the entropy, in a very similar sense as $\hat{p}_{i}$ is an
estimator of $p_{i}$. However, there is an important difference
stemming from the non-linear nature of the entropy functional.
The frequencies $\hat{p}_{i}$ are {\it unbiased} estimators of the
probabilities, i.e., their expectation value $\langle
\hat{p}_{i}\rangle $
(where $\langle \cdot \rangle$ stands for ensemble averages)
 coincides with the true value of the estimated
quantity:
\begin{equation}
\begin{array}{rl}
\langle \hat{p}_{i}\rangle =&\dfrac{\langle n_{i}\rangle }{N}=p_{i}.
\end{array}
\label{p_expect}
\end{equation}
In other words, the frequencies $\hat{p}_{i}$ approximate the
probabilities $p_{i}$ with certain statistical error ({\it variance})
but without any systematic error ({\it bias}).  Contrarily, {\it
  na\"ive} entropy estimators, such as $\hat{H}_{S}^{naive}$, in which
the $p_{i}$ are simply replaced by $n_{i}/N$ are always biased, i.e.
they deviate from the true value of the entropy not only statistically
but also systematically.  Actually, defining an error variable
$\epsilon_{i}=(\hat{p_{i}} - p_{i})/p_{i}$, and replacing
$p_{i}$ in Eq.(1) by its value in terms of $\epsilon_i$ and
$\hat{p_i}$, it is straightforward to verify that the bias, up to
leading order, is $-\frac{M-1}{2N}$, which is a significant error for
small $N$ and vanishes only as $N\rightarrow \infty$ \cite{Roulston}.
A similar bias, owing in general to the nonlinearity of the entropy
functional, appears also for the R\'enyi and Tsallis entropies.

Therefore, the question arises whether it is possible to find improved
estimators which reduce either the bias or the variance of the
estimate.  More generally, the problem can be formulated as follows.
Given an arbitrary entropy functional of the form:

\begin{equation}
\begin{array}{rl}
  H =&F\left[ \displaystyle \sum_{i=1}^{M}h(p_{i})\right] 
\end{array}
\label{H_generic}
\end{equation}
(where $F$ is a generic function) 
we want to find an estimator
\begin{equation}
\begin{array}{rl}
  \hat{H} =&F\left[ \displaystyle \sum_{i=1}^{M}\chi_{n_{i}}\right]
\end{array}
\label{H_est_generic}
\end{equation}
such that the bias
\begin{equation}
\begin{array}{rl}
\Delta=&\langle \hat{H}\rangle -H
\end{array}
\label{H_bias}
\end{equation}
or the mean squared deviation (the statistical error)
\begin{equation}
  \begin{array}{rl}
    \sigma^{2}=&\left\langle \left(\hat{H}-\langle \hat{H}\rangle\right)^{2}\right\rangle
\end{array}
\label{H_var}
\end{equation}
or a combination of both are as small as possible. At the very end of
such a calculation the estimator is defined by $N +1$ real numbers
$\chi_{n_{i}}$ \cite{note2}, which depend on the sample size $N$.  For
example, the na{\"i}ve estimator for the Shannon entropy would be
given in terms of
\begin{equation}
\begin{array}{rl}
\chi_{n_{i}}^{naive}\;=&-\dfrac{n_{i}}{N}\textnormal{ln}\left(\dfrac{n_{i}}{N}\right).
\end{array}
\label{shannon_est_naive}
\end{equation}
The search for improved estimators has a long history. To the best of
our knowledge,
the first to address
 this question was Miller in
1955 \cite{Miller}, who suggested a correction to reduce the bias of
the estimate of Shannon entropy, given by:
\begin{equation}
\begin{array}{rl}
  \chi_{n_{i}}^{Miller}\;=&-\dfrac{n_{i}}{N}\textnormal{ln}\left(
    \dfrac{n_{i}}{N}\right)+\dfrac{1}{2N}.
\end{array}
\label{shannon_est_miller}
\end{equation}
The correction exactly compensates the leading order of the bias, as
reported above.  In this case the remaining bias vanishes as $1/N^{2}$
as $N\rightarrow \infty$.  This result was improved by Harris in 1975
\cite{Harris}, who calculated the next-leading order correction.
However, his estimator depends explicitly on the (unknown)
probabilities $p_{i}$, so that its practical importance is limited.

In another pioneering paper, Grassberger, elaborating upon previous
work by Herzel~\cite{Herzel}, proposed an estimator which provides
further improvement and gives a very good compromise between bias and
statistical error \cite{Grass1}. For the Shannon entropy his estimator
is given by
\begin{equation}
\chi_{n_{i}}^{Grassberger}=\dfrac{n_{i}}{N} \left(
\ln N -\psi(n_i) - \frac{\left( -1\right)^{n_{i}}}{n_{i}(n_{i}+1)} \right),
\label{shannon_est_grass}
\end{equation}
where $\psi(x)$ is the derivative of the logarithm of the Gamma
function, valid for all $i$ with $n_i > 0$. According to
\cite{Grass1}, the function $\psi(x)$ can be approximated by
\begin{equation}
\psi(n_i) \approx \ln x-\frac{1}{2x}
\end{equation}
for large $x$, giving
\begin{equation}
\begin{array}{rl}
\chi_{n_{i}}^{Grassberger}\approx &-\dfrac{n_{i}}{N}\textnormal{ln}\left(\dfrac{n_{i}}{N}
\right)+\dfrac{1}{2N}-\dfrac{\left( -1\right)^{n_{i}}}{N(n_{i}+1)}
\end{array}
\label{shannon_est_grass2}
\end{equation}
This method can be generalized to $q$-deformed entropies.

More recently, a further improvement for the Shannon case has been
suggested by Grassberger \cite{Grass2}
\begin{equation} 
\chi_{n_{i}}^{GS}=\dfrac{n_{i}}{N}\left[
 \psi(N)- \psi(n_i) - ( -1)^{n_{i}} \int_0^1  \frac{t^{n_i-1}}{1+t} dt \right].
\label{new}
\end{equation} This estimator can be recast (see Eqs. (28),(29),(35) of Ref.~\cite{Grass2})  as 
\begin{equation}
\chi_{n_{i}}^{GS}=\frac{n_i}{N}\Bigl(\ln N-G_{n_i}\Bigr)\,,
\end{equation}
where the $G_n$ satisfy the recurrence relation
\begin{eqnarray}
G_1 &=& -\gamma-\ln 2 \\
G_2 &=& 2-\gamma-\ln 2 \\
G_{2n+1} &=& G_{2n} \\
G_{2n+2} &=& G_{2n}+2/(2n+1)
\end{eqnarray}
with $\gamma=-\psi(1)$. This estimator constitutes the state of the art for Shannon
entropies, but unfortunately, it cannot be straightforwardly extended
to more general q-deformed entropy functionals, for which
\cite{Grass1} remains the best available option. These results were further
generalized by Sch{\"u}rmann~\cite{Schuermann} with different balances
between statistical and systematic errors.

It should be emphasized that an ideal estimator does not exist,
instead the choice of the estimator depends on the structure of data
to be analyzed \cite{SG}. For example, the above discussed estimators
\cite{Grass1,Grass2} work satisfactorily if the probabilities $p_{i}$
are sufficiently small. This is the case in many applications of
statistical physics, where the number of possible states, $M$, in an
ensemble is usually extremely large so that the probability~$p_{i}$
for an individual state $i$ is very small. On the other hand, this
assumption does not always hold for empirical data sets such as
digital data streams and DNA sequences.

The performance of the estimators worsens as
the values of $p_{i}$ get larger. This is due to the following
reason: the numbers $n_{i}$, which count how often the value $x_{i}$
appears in the data set, are generically distributed as binomials,
i.e. the probability $P_{n_{i}}$ to find the value $n_{i}$ is given
by:
\begin{equation}
\begin{array}{rl}
P_{n_i}({p_i}) =&\dbinom{N}{n_{i}}p_{i}^{n_{i}}(1-p_{i})^{N-n_{i}}
\end{array}
\label{binomial_pdf}
\end{equation}
where $\dbinom{N}{n_{i}}=\dfrac{N!}{n_{i}!(N-n_{i})!}$ are binomial
coefficients. For $p_{i}\ll1$ this can be approximated by a Poisson
distribution, which is the basis for the derivation of
Eq.~(\ref{shannon_est_grass}). For large values
$p_{i}$, however, this assumption is no longer justified and this
results in large fluctuations (even if the bias remains small).

It is important to note that it is not possible to design an
estimator that minimizes both the bias and the variance to arbitrarily
small values. The existing studies have shown that there is always a
delicate tradeoff between the two types of errors.  For example,
minimizing the bias usually comes at the expense of the variance,
which increases significantly. Moreover, it can be proved that neither
the variance nor the bias can be reduced to zero for finite $N$
\cite{Paninski}. Therefore, it is necessary to study estimators
with different balances between systematic and statistical errors,
as it was done e.g. in the work by Sch{\"u}rmann~\cite{Schuermann}.

In the present work we introduce two estimators, which can be used to
measure any of the entropy functionals discussed above. Both of them
are specifically designed for short data series where the
probabilities $p_{i}$ take (in general) non-small values.  The first
one reduces the bias as much as possible at the expense of the
variance, and is mostly of academic interest and discussed only for
illustration purposes. The second one seeks for a robust compromise
between minimizing bias and variance together, is very easy to
implement numerically, and has a broad potential range of
applicability. The estimator itself can be improved by adapting various
of its elements to each specific problem.

\section{Low-bias estimator}

The starting point is the observation that the entropy $H$ and its
estimators $\hat{H}$ in Eq.~(\ref{H_generic}) and
Eq.~(\ref{H_est_generic}) involve sums over all possible values of
the data set. Therefore,
as the bias
 can be minimized by minimizing the errors of each
summand, the problem can be reduced to minimize 
\begin{equation}
\begin{array}{rl}
  \delta(p_{i})=&\langle \chi_{n_{i}}\rangle -h(p_{i})=\left( 
    \displaystyle \sum_{n_{i}=0}^{N}P_{n_{i}}(p_{i})\chi_{n_{i}}\right)-h(p_{i})
\end{array}
\label{box_bias2}
\end{equation}
over a broad range of $p_{i}$ as much as possible.

A theorem by Paninski \cite{Paninski} states that it is impossible to
reduce the bias to zero for all $p_{i}\in [0, 1]$ since an estimator
is always a finite polynomial in $p_{i}$ while the true entropy is
usually not a polynomial. However, it is possible to let the bias
vanish at 
$N+1$
 points $p_{i}$ in the interval $[0, 1]$ because the
determination of the different $\chi_{n_i}$ requires 
$N+1$
 independent equations.

For the sake of illustration, let us choose here
equidistant points $p_{j} = j/N$, with $j=0,1,\hdots,N$. In general,
other choices, more appropriate to each specific case, should be
employed. The resulting set of linear equations reads:
\begin{equation}
  \begin{array}{rl}
\delta(j/N)=0\Longrightarrow&\displaystyle \sum_{n_{i}=0}^{N}P_{n_{i}}(j/N)\chi_{n_{i}}=h(j/N),
\hspace{2cm} j=0,1,\hdots, N.
\end{array}
\label{low_bias}
\end{equation}

\begin{figure}
 \centering\includegraphics[width=90mm]{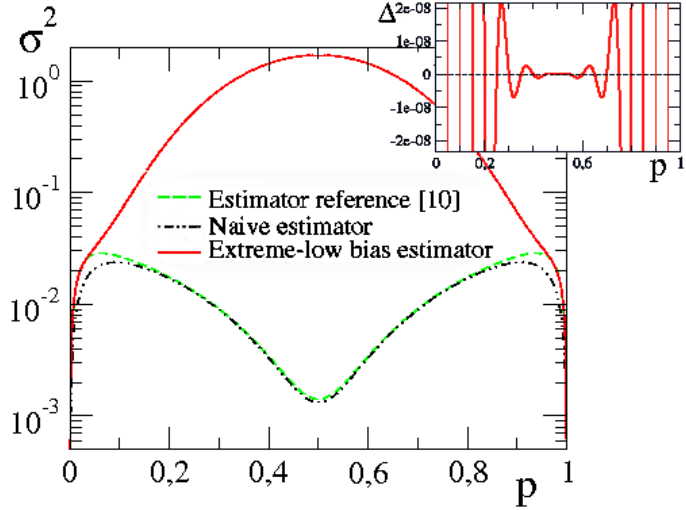}
 \caption{\label{Fig1} Fluctuations, $\sigma^2$, as defined by
   Eq.~(\ref{H_var}), for three different Shannon entropy estimates
   (the na\"ive one, the improved estimator introduced in
   \cite{Grass2}, and the low-bias estimator defined in this paper)
   for a binary sequence ($M=2$) of length $N=20$.  Inset: Bias,
   $\Delta$, as defined by Eq.~(\ref{H_bias}), for the low-bias
   estimator showing the N+1 vanishing points with amplitude
   oscillations.}
\end{figure}
\noindent
Introducing the notation $h_{j}=h(j/N)$ and
$P_{j,n_{i}}=P_{n_{i}}(j/N)$ 
this last expression takes the form:
\begin{equation}
\begin{array}{rl}
\displaystyle \sum_{n_{i}=0}^{N}P_{j,n_{i}}\chi_{n_{i}}=&h_{j},
\hspace{2cm} j=0,1,\hdots, N
\end{array}
\label{low_bias2}
\end{equation}
or, in short, ${\bf P} \overrightarrow{\chi} = \overrightarrow{h}$,
where ${\bf P}$ is the so-called {\it multinomial matrix}
\cite{multi_examples}.  To find the solution $\overrightarrow{\chi} =
{\bf P}^{-1} \overrightarrow{h}$, the matrix:
\begin{equation}
\begin{array}{rl}
  P_{j,n_{i}} =&\dbinom{N}{n_{i}}p_{j}^{n_{i}}(1-p_{j})^{N-n_{i}}=
  \dbinom{N}{n_{i}}\left(\dfrac{j}{N}\right)^{n_{i}}\left(1-\dfrac{j}{N}\right)^{N-n_{i}},
\end{array}
\label{multinomial_matrix}
\end{equation}
whose 
elements are
 binomial distributions, has to be inverted.  For
small $N$ this inversion is most easily done numerically. However, we
were also able to invert the matrix analytically, leading us to the
closed form \cite{future}:
\begin{equation}
\begin{array}{cc}
  \hat{P}_{i,j}^{-1}=&\displaystyle \sum_{k=0}^{N} \sum_{l=0}^{N} 
  \dbinom{i}{k}\dbinom{l}{j}\dfrac{N^{k}k!(N-k)!}{N!}\dfrac{(-1)^{l+j}}{l!}
s(l,k)
\end{array}
\label{multi_node_inv}
\end{equation}
where $s(l,k)$ denotes the Stirling numbers of the first kind \cite{AS}.
 Having
inverted the matrix, the numbers $\chi_{n_{i}}$ determining the
estimators can be computed for any given entropy functional by a
simple matrix multiplication.

Figure 1 illustrates a comparison for the Shannon case between the
low-bias estimator and other well-known ones for the simple example of
a binary sequence of $N=20$ bits $x = 0, 1$ (i.e.  $M=2$), where the
value $1$ appears with probability $p$ and $0$ with probability $1-p$.
The bias of the low-bias estimator vanishes exactly only at values of
$p$ multiples of $1/20$, and takes small values in between (see inset
of Fig.1).  On the other hand, the fluctuations for both the na\"ive
estimator and the one in \cite{Grass2} remain bounded, while they
diverge for the low-bias case (Fig.1 ). This unbounded growing of
statistical fluctuations makes the low-bias estimator useless for
practical purposes.

\section{Balanced Estimator}

Aiming at solving the previously illustrated problem with uncontrolled
statistical fluctuations, in this section we introduce a new {\it
  balanced estimator} designed to minimize simultaneously both the
bias and the variance over a wide range of probabilities.  This is of
relevance for analyzing small data sets where statistical fluctuations
are typically large and a compromise with minimizing the bias is
required.

As before, ignoring correlations between the $n_{i}$ both bias and
statistical errors can be optimized by minimizing the errors of the
summands in their corresponding expressions. Therefore, the problem
can be reduced to minimize the bias for each state
\begin{equation}
\begin{array}{rl}
\delta(p_{i})=&\langle \chi_{n_{i}}\rangle -h(p_{i})
\end{array}
\label{box_bias}
\end{equation}
and the variance within such a state
\begin{equation}
\begin{array}{rl}
      \sigma^{2}(p_{i})=&\left\langle \left( \chi_{n_{i}}-
      \langle\chi_{n_{i}}\rangle \right)^{2}\right\rangle
\end{array}
\label{box_var}
\end{equation}
over a broad range of $p_{i}\in [0, 1]$, where $n_{i}\in {0,
  1,\hdots,N}$ is binomially distributed.
Since we are interested in a balanced compromise error, it is natural
to minimize the squared sum:
\begin{equation}
  \Phi^{2}(p_{i})= \delta^{2} (p_{i})+\sigma^{2} (p_{i}).
\label{square}
\end{equation}
This quantity measures the total error for a particular value of
$p_{i}$. Therefore, the average error over the whole range of
$p_{i}\in [0, 1]$ is given by:
\begin{equation}
\begin{array}{rl}
\overline{\Phi^{2}_{i}}=&\displaystyle\int^{1}_{0}dp_{i}w(p_{i})\Phi^{2}(p_{i})
\end{array}
\label{av_error}
\end{equation}
where $w(p_{i})$ is a suitable weight function that should be
determined for each specific problem.

We discuss explicitly here the simplest case $w(p_{i})\equiv1$
(obviously, any extra knowledge of the probability values should lead
to a non-trivial distribution of weights, resulting in improved
results).  Inserting Eq.  (\ref{box_bias}) and Eq. (\ref{box_var})
into Eq. (\ref{av_error}), the average error is given by:
\begin{equation}
\begin{array}{rl}
  \overline{\Phi^{2}_{i}}=&\displaystyle\int^{1}_{0}dp_{i}\left[\left( \displaystyle 
      \sum_{n_{i}=0}^{N}P_{n_{i}}(p_{i})\chi_{n_{i}}^{2}\right) +h^{2}(p_{i})-2h(p_{i})
    \left( \displaystyle \sum_{n_{i}=0}^{N}P_{n_{i}}(p_{i})\chi_{n_{i}}\right)\right].
\end{array}
\label{av_error2}
\end{equation}

Now, we want to determine the numbers $\chi_{n_{i}}$ in such a way
that
the error given by Eq.\ref{av_error2}
 is minimized.
Before proceeding, let us make it clear that instead of minimizing the
mean-square error for each of the possible states ($i=1,...,M$) one
could also minimize the total mean-square error defined using
Eq.(\ref{H_bias}) and (\ref{H_var}) rather than Eq.  (\ref{box_bias})
and (\ref{box_var})
 to take into account correlations between boxes which, in general, will improve the final result.
 For example, for binary sequences this can be
easily done, and leads to the same result as reported on what follows
\cite{future}.

As a necessary condition, the partial derivatives:
\begin{equation}
\begin{array}{rl}
\dfrac{\partial}{\partial\chi_{n_{i}}}\overline{\Phi^{2}_{i}}=&0
\end{array}
\label{min_condition}
\end{equation}
have to vanish, i.e.:
\begin{equation}
\begin{array}{rl}
  2 \displaystyle\int^{1}_{0}dp_{i}P_{n_{i}}(p_{i})\left[\chi_{n_{i}}-h(p_{i})\right]  =&0
\end{array}
\label{min_condition2}
\end{equation}
for all $n_{i} = 0, 1, \hdots,N$. Therefore, the balanced estimator is
defined by the numbers:
\begin{equation}
\begin{array}{rl}
  \chi_{n_{i}}^{bal}=&\dfrac{\displaystyle\int^{1}_{0}dp_{i}P_{n_{i}}(p_{i})h(p_{i})}
  {\displaystyle\int^{1}_{0}dp_{i}P_{n_{i}}(p_{i})}=
  (N+1)\displaystyle\int^{1}_{0}dp_{i}P_{n_{i}}(p_{i})h(p_{i}).
\end{array}
\label{balanced_est_def}
\end{equation}
where we have explicitly integrated over $p_i$ the binomial
distribution.

In the Shannon case, where $h(p_{i})=-p_{i}\textnormal{ln}(p_{i})$,
the integration can be explicitely carried out, leading to
\cite{note3}:
\begin{equation}
\begin{array}{rl}
\chi_{n_{i}}=&\dfrac{n_{i}+1}{N+2}\displaystyle\sum_{j=n_{i}+2}^{N+2}\dfrac{1}{j}
\end{array}
\label{balanced_est_shannon}
\end{equation}
so that the final result for the balanced estimator of Shannon entropy
is given by:
\begin{equation}
\begin{array}{rl}
\hat{H}_{S}^{bal}=&\dfrac{1}{N+2}\displaystyle\sum_{i=1}^{M}\left[(n_{i}+1)
\displaystyle\sum_{j=n_{i}+2}^{N+2}\dfrac{1}{j}\right].
\end{array}
\label{balance_shannon}
\end{equation}
Similarly, it is possible to compute $\chi_{n{i}}$ for a power
$h(p_{i}) = p_{i}^{q}$, which is the basis for all $q$-deformed
entropies:
\begin{equation}
\begin{array}{rl}
\chi_{n_{i}}(q)=&\dfrac{\Gamma(N+2)\Gamma(n_{i}+1+q)}{\Gamma(N+2+q)\Gamma(n_{i}+1)}
\end{array}
\label{balanced_est_qdef}
\end{equation}
The balanced estimators for R\'enyi \cite{note4} and Tsallis entropy
are then given respectively  by:
\begin{equation}
\hat{H}_{R}^{bal}(q)=\dfrac{1}{1-q}\textnormal{ln}\left[\displaystyle
\sum_{i=1}^{M}
\chi_{n_{i}}(q)
\right], ~~ 
\label{balance_renyi}
\end{equation}
and
\begin{equation}
  \hat{H}_{T}^{bal}(q)=\dfrac{1}{q-1}\left[1-\displaystyle\sum_{i=1}^{M}
\chi_{n_{i}}(q)
\right].
\label{balance_tsallis}
\end{equation}
\begin{figure}[t]
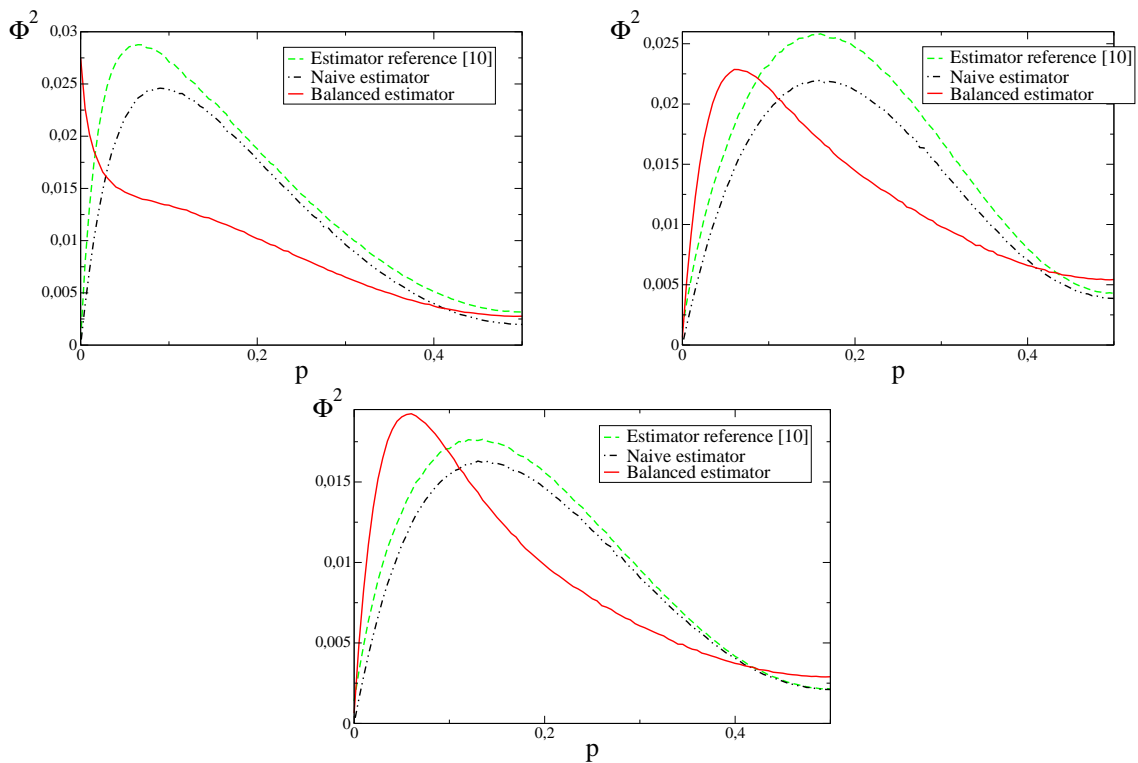

\begin{center}
\includegraphics[height=5.0cm]{Fig2a.eps}\hspace{1cm}
\includegraphics[height=5.0cm]{Fig2b.eps}
\includegraphics[height=5.0cm]{Fig2c.eps}
\end{center}
\caption{\footnotesize{ Mean squared error $\Phi^{2}=\langle
    \left(\hat{H}-H\right)^{2}\rangle$ of different entropy estimators
    (Upper row: Shannon (left); R\'enyi with $q=1.5$ (right); Lower row: Tsallis with $q=1.5$)
    for a binary sequence
    of $N = 20$, as a function of $p$. The set of possible values is
    $\left\lbrace x_{1}=1, x_{2}=0 \right\rbrace$ and the
    probabilities, $\left\lbrace p_{1}=p, p_{2}=1-p \right\rbrace$,
    respectively.  }}
\label{Fig2}
\end{figure}

To illustrate the performance of these estimators, let us consider
again a binary sequence of $N$ bits $x = 0, 1$ (i.e.  $M=2$) occurring
with probabilities $1-p$ and $p$ respectively. In Fig. \ref{Fig2} we
plot the mean squared deviation $\Phi^{2}= \langle \left(
  \hat{H}-H\right)^{2}\rangle$ of various estimators from the true
value of the Shannon as well as the R\'enyi entropy as a function of
$p$. For such a short bit sequence, the performance of the
Grassberger's estimator {\it using the parameter $\Phi^2$}, is even
worse than the na\"ive one.  This is not surprising since
Grassberger's estimator is designed for small probabilities
$p_{i}\ll1$, while in the present example one of the probabilities $p$
or $1 - p$ is always large
 and thus the estimator is affected by large fluctuations.
 The balanced estimator, however, reduces
 mean squared error considerably over an extended range of $p$ while
 for small $p$ and $0.4<p<0.6$ it fails.  
Similar plots can be obtained for the Tsallis entropy.

The advantage of the balanced estimator compared to standard ones
decreases with increasing $N$. One of the reasons is the circumstance
that the fluctuations of the estimator are basically determined by the
randomness of the $n_{i}$ and, therefore, are difficult to reduce.

\section{Conclusions}

We have designed a new ``balanced estimator'' for different entropy
functionals (Shannon, R\'enyi, and Tsallis) specially adequate for the
analysis of small data sets where the possible states appear with
not-too-small probabilities.  To construct it, first we have
illustrated a known result establishing that systematic errors (bias)
and statistical errors cannot both be simultaneously reduced to
arbitrarily small values when constructing an estimator for a limited
data set. In particular, we have designed a low-bias estimator and
highlighted that it leads to uncontrolled statistical fluctuations.
This hinders the practical usefulness of such a low-bias estimator.

On the other hand, we have designed a new estimator that constitutes a
good compromise between minimizing the bias and keeping controlled
statistical fluctuations.  We have illustrated how this balanced
estimator outperforms (in reducing simultaneously bias and
fluctuations) previously available ones in special situations the data
sets are sufficiently small and the probabilities are not too small.
Obviously situations such as in Fig. 2 are the `worst case' 
for estimators like~(\ref{shannon_est_grass}) 
and~(\ref{new}) which were designed to be efficient for large $M$.
If any of these conditions is not fulfilled
Grassberger's and Sch\"urmann's estimator remains the best choice.

The balanced method fills a gap in the list of existing entropy
estimators, is easy to implement for Shannon, R\'enyi and Tsallis
entropy functional and therefore its potential range applicability is
very large, specially in analyses of short biological (DNA, genes,
etc.)  data series.

The balanced estimator proposed here is simple but by no means 
`optimal' for two reasons. First, we made no effort to optimize
the location of the mesh points $p_j$, which for simplicity are
assumed to be equidistant. Moreover, we did not optimize the weights
$w(p_j)$ towards a Bayesian estimate, as e.g. attempted by Wolpert
and Wolf~\cite{Wolpert}. Further effort in this direction would 
be desirable.

\vspace{0.25cm} We acknowledge financial support from the Spanish
Ministerio de Educaci\'on y Ciencia (FIS2005-00791) and Junta de
Andaluc{\'\i}a (FQM-165).


\begin{thebibliography}{10}
\bibitem{note2} This is so because the estimator $\chi_{n_{i}}$
  depends only on $n_{i}$, which can take $N+1$ possible values.
\bibitem{note4} Here the influence of the non-linearity of the
  logarithm on statistical averages is neglected.
\end{thebibliography}
\end{document}